\documentclass[runningheads]{llncs}
\usepackage{pgfplots}
\usepackage[linesnumbered,ruled, vlined]{algorithm2e}
\usepackage{setspace}
\usepackage{hyperref}
\usepackage{graphicx}
\usepackage{verbatim}
\pgfplotsset{compat=1.9}
\usepgfplotslibrary{groupplots}
\usetikzlibrary{matrix}
\SetAlFnt{\small\sf}

\begin{document}
\title{Internal Data Imputation \\in
Data Warehouse Dimensions}
\author{Yuzhao Yang\inst{1} \and Fatma Abdelhédi \inst{3} \and
Jérôme Darmont \inst{2}\and
Franck Ravat\inst{1}\and  Olivier Teste\inst{1}}
\authorrunning{Y. Yang et al.}
\institute{IRIT-CNRS (UMR 5505), 
    Université de Toulouse, France \\\email{\{Yuzhao.Yang, Franck.Ravat, Olivier.Teste\}@irit.fr} \and
 Université de Lyon, Lyon 2, UR ERIC, France \\
    %\\5 avenue Pierre Mendès France, F69676 Bron Cedex, France
\email{jerome.darmont@univ-lyon2.fr}\and
CBI$^2$ – TRIMANE, Paris, France\\
\email{Fatma.Abdelhedi@trimane.fr}}
\maketitle  

\begin{abstract}
Missing data occur commonly in data warehouses and may generate data usefulness problems. Thus, it is essential to address missing data to carry out a better analysis. There exists data imputation methods for missing data in fact tables, but not for dimension tables. Hence, we propose in this paper a data imputation method for data warehouse dimensions that is based on existing data and takes both intra- and inter-dimension relationships into account.
\end{abstract}
\section{Introduction}
%\subsubsection{Context}
Data warehouses (DWs) are widely used in companies and organizations  to help building decision support systems. Data in DWs are usually modeled in a multidimensional way,  which helps users consult and analyze aggregated data with On-Line Analytical Processing (OLAP). In a DW, there are non-NULL constraints on keys, but not always on the other attributes, so there may be missing data. Missing data may come from the DW's sources (operational data sources, data of other DWs) if not treated during the Extract-Transform-Load (ETL) process. We classify missing data in the DWs into factual missing data and dimensional missing data with respect to their occurrence in fact and dimension tables. Factual missing data are usually quantitative, making analysis results incomplete and preventing users from getting reliable aggregates. Dimensional missing data are usually qualitative, making aggregated data incomplete and making it hard to analyse them with respect to hierarchy levels. Therefore, it is significant to complete the missing data for the sake of a better data analysis.
%\vspace{-10pt}
%\subsubsection{Related work}

Data imputation is the process of filling in missing data by plausible values based on information available in the dataset \cite{Li04}. %Eliminating the impact of missing data in DWs, it is a significant task to conduct the data imputation for the sake of a better data analysis.
%There exists imputation methods that replace factual missing data 
%The existing work concerning the 
Imputation of missing data focuses on factual data, with statistic-based \cite{wu02}, K-Nearest Neighbour (KNN)-based \cite{Ribeiro11}, linear programming-based \cite{bimont20} and hybrid (KNN and constraint programming) \cite{Amanzougarene14} methods.
There is no research about dimensional missing data. However, dimensional data are mostly qualitative and there are methods for qualitative data imputation. Some methods replace missing values through business rules \cite{Fan10,Son15} or association rules \cite{Wu04,Shen07}. Yet business rules are not always available in practice, and association rules require to define support and confidence thresholds, which is not always easy. External sources can also be employed, e.g., through crowdsourcing \cite{Lod13} or taking advantage of Web information \cite{Yak12}. % like web pages, web lists and web tables.
%\vspace{-10pt}
Yet, suitable external sources may be difficult to find. 
%It can be seen that there is no research in regard to completing the dimensional missing data in the DWs. Concerning the existing methods for qualitative missing data imputation, for the rule based methods, the business rules are not always available, association rules require the user to define a support and confidence. The external source based methods need the corresponding sources which may be difficult to be found. 
Eventually, data imputation in DWs should consider the different structural elements in an OLAP systems, such as dimensions and hierarchies. As a result, we propose in this article an internal, i.e., based on existing data, data imputation method for dimensional missing data in DWs, by considering inter- and intra-dimension relationships.

The rest of the paper is organized as follow. In Section~\ref{sec:prel}, we formalize the OLAP model. In Section~\ref{sec:imputation}, we detail our imputation method and provide the corresponding algorithms. In Section~\ref{sec:xp}, we validate our proposal through a series of experiments. Finally, in Section~\ref{sec:conclu}, we conclude the paper.

%This motivate us to propose our imputation method for the dimensional data.

%However, to the best of our knowledge, there is no specific imputation approach devoted to dimensional missing data. Moreover, data imputation in DWs should consider the different structural elements in an OLAP system, such as dimensions and hierarchies. %and dimensional data are usually qualitative data. 
%As a result, we propose in this article an internal, i.e., based on existing data, data imputation method for dimensional missing data in DWs, by considering inter- and intra-dimension relationships.% Internal means the method is based on the existing data.

\section{Preliminaries}
\label{sec:prel}

We introduce here the multidimensional DW concepts and notations used in this paper~\cite{Ravat08}. 

\begin{definition}
A data warehouse, denoted $DW$, is defined as ($N^{DW}, F^{DW},$ $D^{DW},$ $Star^{DW}$), where $N^{DW}$ is the data warehouse's name, $F^{DW} = \{F_1^{DW}, ...,$  $F_m^{DW}\}$ is a set of facts, $D^{DW} = \{D_1^{DW}, ... ,D_n^{DW}\}$ is a set of dimensions and $Star^{DW}:F^{DW} \rightarrow D^{DW}$ is a mapping associating each fact to its linked dimensions.
\end{definition}
\begin{definition}
A dimension, denoted $D \in D^{C}$, is defined as ($N^D, A^D, H^D, I^D$), where $N^D$ is the dimension's name, $A^D = \{ a^D_1,...,a^D_u\} \cup \{id^D\}$ is a set of attributes, where $id^D$ represents the dimension's identifier. $H^D = \{H^D_1,...,H^D_v\}$ is a set of hierarchies. $I^D = \{i^D_1,...,i^D_e\}$ is a set of dimension instances. The value of instance $i^D_e$ for attribute $a^D_u$ is denoted as $i^D_e.a^D_u$.
\end{definition}

\begin{definition}
A hierarchy of dimension $D$, denoted $H \in H^D$, is defined as $(N^{H}, Param^{H})$, where $N^{H}$ is the hierarchy's name. $Param^{H} = <id^D, p^H_2, ..., p^H_v$ $>$ is an ordered set of dimension attributes, called parameters, which set granularity levels along the dimensions: $\forall k \in [1...v], p^H_k \in A^D$. The case where $p^H_1$ rolls up to $p^H_2$ in $H$ is denoted by $p^H_1 \preceq_H p^H_2$. $Weak^{H}$ = $Param^{H} \rightarrow (A^D - Param^{H})$ is a mapping possibly associating each parameter with one or several weak attributes, which are also dimension attributes providing additional information. $Weak^{H}[p^H_x] = \{w^{p^H_x}_1\,...,w^{p^H_x}_y\}$ is the weak attribute set for parameter $p^H_x$. 
\end{definition}

\section{Internal Data Imputation for Dimensions}
\label{sec:imputation}

Internal data imputation consists in replacing missing data in dimensions with the aid of existing data. Existing data imputation is convincible because we use accurate data and not predictions or otherwise computed values. Imputation can be achieved through intra- and inter-dimensional relationships. Let us introduce these two types of data imputation.
%\vspace{-10pt}

\subsubsection{Intra-dimensional Imputation}

Intra-dimension imputation relies on data from the same dimension. % to complete the missing data. (Redondant avec imputation)
There are %relationships of 
indeed functional dependencies between attributes in the same hierarchy. If an attribute is a parameter, its values depend on the values of lower-granularity parameters. %We take an example of the geographical hierarchy which often comes out in the DWs, if we have $city \rightarrow country \rightarrow continent$. The city Paris belongs always to the country France, if for an instance whose value of city is Paris, even if the country value is missing, we know that it should be France.
Our intra-dimension imputation method is presented in Algorithm~\ref{algo:duplicate1}. We first check each parameter in hierarchies of the DW. If there exists missing data for this parameter (Lines 1-2), we search for an instance with value in a lower-granularity parameter and whose value exists (Lines 3-4). Then, we can then fill in the missing data with this value (Line~5). 
%\vspace{-10pt}

\begin{algorithm}[hbt]
\setstretch{1.2}
\DontPrintSemicolon
\For{each $p^H_v \in Param^H$, where $H \in H^D, D \in D^{DW}$}{
    \For{each $i^D_e \in I^D,$ where $i^D_e.p^H_v$ is null}{
            \While{$p^H_{v_2} \in Param^H \land p^H_{v_2} \preceq_H p^H_v$}{
        \If{$\exists i^D_{e_2} \in I^D, i^D_{e_2}.p^H_{v_2} = i^D_e.p^H_{v_2} \land i^D_{e_2}.p^H_v$ is not null}{
            $i^D_{e}.p^H_v \gets i^D_{e_2}.p^H_{v_2}$}}}

    \For{each $i^D_{e_3} \in I^D$, where $i^D_{e_3}.w^{p^H_v}_y$ is null, $w^{p^H_v}_y \in Weak^H[p^H_v]$}{
    \While{$p^H_{v_3} \in Param^H \land (p^H_{v_3} \preceq_H p^H_v \lor p^H_{v_3} = p^H_v)$)}{
        \If{$\exists i^D_{e_4} \in I^D, i^D_{e_4}.p^H_{v_3} = i^D_{e_3}.p^H_{v_3} \land i^D_{e_4}.p^H_v$ is not null}{
            $i^D_{e_3}.w^{p^H_v}_y \gets i^D_{e_4}.w^{p^H_v}_y$}}}
}
\caption{Intra-dimension Imputation}
\label{algo:duplicate1}
\end{algorithm}

The value of a weak attribute depends on the values of its parameter. Then, for each weak attribute of the parameter we check, if there are missing data (Line 6), we search for the instance that has the same value of its parameter or a lower-granularity parameter whose value exists (Lines 7-8). The missing weak attribute data can then be supplied by this value (Line 9). It is important to note that, since the parameter sets of hierarchy are ordered sets, checking parameters is sequential (from the lowest-granularity to the highest-granularity parameter). This ensures that  imputation is maximal, as the value of a higher-granularity parameter depends on its lower-granularity parameters. %If we complete the values of a higher parameter before the imputation of its lower parameters, there may be some missing data of the lower parameters which are not yet completed but which can be used to complete the higher parameters so that the imputation is not maximal.

%\vspace{-10pt}

%\vspace{-20pt}

\subsubsection{Inter-dimensional Imputation}

In a DW, there may  be attributes that are common to different dimensions. Therefore, we can replace missing data with such inter-dimensional common attributes. The main idea of inter-dimension imputation is similar to intra-dimension imputation's, except that instead of searching for parameters in the same hierarchy, we search for common parameters of hierarchies in other dimensions (Algorithm~\ref{algo:duplicate2}, Lines 3-4 and 9-10). When performing the imputation of weak attributes, we must make sure that, in the searched dimension, the searched parameter is semantically identical with the parameter of the weak attribute to be completed; and that it bears a semantically identical weak attribute (Lines 10-11). We say ``semantically identical'' because in a DW, common attributes may be presented differently in different dimensions. Since in a DW, the designer would normally not use two vocabularies to describe a same entity, but may use the different prefixes or suffixes to distinguish the same entity in different dimensions, we must therefore use string similarity to match attribute names. 
%\vspace{-0pt}
\begin{algorithm}
\setstretch{1.2}
\DontPrintSemicolon
\For{each $p^H_v \in Param^H$, where $H \in H^D, D \in D^{DW}$}{
  \For{each $i^D_e \in I^D$, where $i^D_e.p^H_v$ is null}{
        \For{each $p^{H_2}_{v_2} \in Param^{H_2}$, where $H_2 \in H^{D_2}$,  $D_2 \in D^{DW} \land D_2 \neq D$}{
            \If{$p^{H_2}_{v_2} \simeq p^H_v$}{
                \While{$p^{H_2}_{v_3} \in Param^{H_2} \land p^{H_2}_{v_3} \preceq_{H_2} p^{H_2}_{v_2}$}{
                    \If{$\exists i^{D_2}_{e_2} \in I^{D_2}, i^{D_2}_{e_2}.p^H_{v_3} = i^D_e.p^H_{v_3} \land i^{D_2}_{e_2}.p^{H_2}_{v_2}$ is not null}{
                        $i^D_e.p^H_v \gets i^{D_2}_{e_2}.p^{H_2}_{v_2}$}}}}
    }
    \For{each $i^D_{e_3} \in I^D$, where $w^{p^H_v}_y \in Weak^H[p^H_v]$, $i^D_{e_3}.w^{p^H_v}_y$ is null}{
    \For{each  $p^{H_3}_{v_4} \in H_3$, where $H_3 \in H^{D_3}$, $D_3 \in D^{DW} \land D_3 \neq D$}{
        \If{$p^{H_3}_{v_4} \simeq p^H_v \land \exists w^{p^{H_3}_{v_4}}_{y_2} \in Weak^{H_3}[p^{H_3}_{v_4}], w^{p^{H_3}_{v_4}}_{y_2} \simeq w^{p^H_v}_y$}{
            \While{$p^{H_3}_{v_4} \in Param^{H_3} \land (p^{H_3}_{v_5} \preceq_{H_3} p^H_v \lor p^{H_3}_{v_5} \simeq p^H_v)$}{
            \If{$\exists i^{D_3}_{e_4} \in I^{D_3}, i^{D_3}_{e_4}.p^{H_3}_{v_5} = i^D_{e_3}.p^{H_3}_{v_5} \land i^{D_3}_{e_4}.w^{p^{H_3}_{v_4}}_{y_2}$ is not null}{$i^D_{e_3}.w^{p^H_v}_y \gets i^{D_3}_{e_4}.w^{p^{H_3}_{v_4}}_{y_2}$}}}}}
  }
 \caption{Inter-dimension Imputation}
\label{algo:duplicate2}
\end{algorithm}

\section{Experimental Assessment}
\label{sec:xp}

We implement our algorithms\footnote{\href{https://github.com/BI4PEOPLE/Internal-Data-Imputationin-Data-Warehouse-Dimensions/}{https://github.com/BI4PEOPLE/Internal-Data-Imputationin-Data-Warehouse-Dimensions/}} and conduct experiments with different datasets. Our code is developed in Python 3.7 and is executed on a Intel(R) Core(TM) i5-10210U 1.60~GHz CPU with a 16~GB RAM. Data are integrated in R-OLAP format with Oracle 11g.

\subsection{Datasets and Experimental Method}
Our experiments are based on one benchmark dataset and three real-world datasets. The % first one is the 
TPC-H benchmark (\textbf{TPCH}) provides a relational schema\footnote{\href{http://tpc.org/tpc\_documents\_current\_versions/pdf/tpc-h\_v2.18.0.pdf}{http://tpc.org/tpc\_documents\_current\_versions/pdf/tpc-h\_v2.18.0.pdf}} with 8 tables and a data generator we use to produce 100~MB of data. The first real-world dataset is a customer-centric dataset (\textbf{GlobalStore}) of a global super store\footnote{\href{https://data.world/vikas-0731/global-super-store}{https://data.world/vikas-0731/global-super-store}}. It contains the order data of different customers and products. The second real world dataset is a regional sale dataset (\textbf{RegionalSales}) storing sales data for a company across US regions\footnote{\href{https://data.world/dataman-udit/us-regional-sales-data}{https://data.world/dataman-udit/us-regional-sales-data}}. The third dataset (\textbf{GeoFrance}) contains information about French cities, departments and regions from the French government open data site\footnote{\href{https://www.data.gouv.fr/fr/datasets/communes-de-france-base-des-codes-postaux/}{https://www.data.gouv.fr/fr/datasets/communes-de-france-base-des-codes-postaux/}}. We create a DW for each real-world dataset.

In our experiments, the parameter of the first granularity level in dimension hierarchies is the primary key of the dimension table. % which cannot be empty. By definition.
Its values are not repetitive, so weak attributes of the first granularity level and  parameters of the second granularity level cannot be completed. Therefore, we generate missing data for parameters from the third granularity level of dimension hierarchies and for weak attributes from the second granularity level. Moreover, we apply different missing rates (1\%, 5\%, 10\%, 20\%, 30\%, 40\% and 50\%). To generate a certain percentage of missing data for an attribute, we sort randomly all the tuples and remove attribute data of the first certain percentage of tuples. For each dataset, we carry out 20 tests and get the average imputation rate, accuracy and runtime. Imputation rate is the number of replaced values divided by the number of missing values. Accuracy is the number of correctly replaced values divided by the number of all replaced values.

\subsection{Intra-dimensional Imputation Experiments}

The datasets \textbf{TPCH}, \textbf{GlobalStore} and \textbf{RegionalSales} are employed in this experiment intra-dimensional imputation experiment. %Mmissing data are generated by the way mentioned in Section\ref{sec:inter}. Repetition
The imputation rate ranges between 61.73\% and 100\%; the accuracy between 97.08\% and 100\%; and missing rate between 1\% and 50\%.
%%\vspace{-10pt} Interdit !

\subsubsection{Imputation Rate}
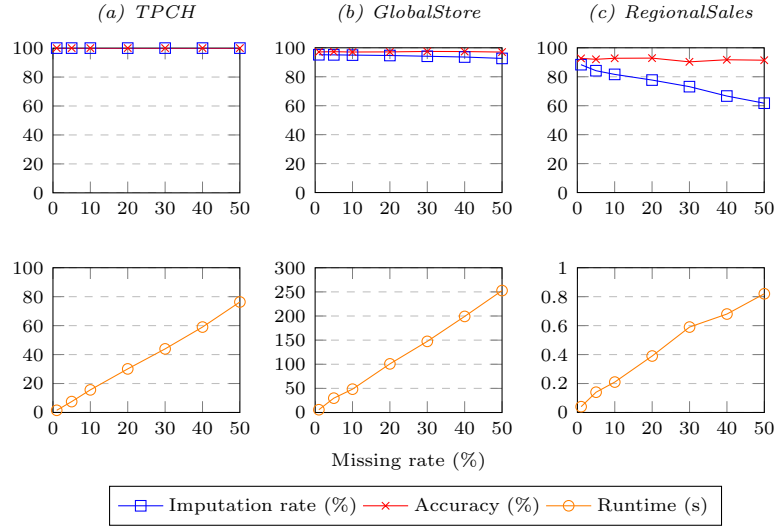
\begin{figure}
\centering
\begin{tikzpicture}
\begin{groupplot}[
   group style={
     group size=3 by 2,
     horizontal sep=1cm},
   width=\linewidth/3,/tikz/font=\scriptsize]
\nextgroupplot[title = {\emph{(a) TPCH}},    
    xmin=0, xmax=50,
    ymin=0, ymax=100,
    xtick={0,10,20,30,40,50},
    ytick={0,20,40,60,80,100},
    ymajorgrids=true,
    grid style=dashed,
    legend style={at={(0.3,-2.3)},anchor=south west, legend columns=3}]
\addplot[color=blue,mark=square]
    coordinates {(1,100)(5,100)(10,100)(20,100)(30,100)(40,100)(50,100)};
\addplot[color=red,mark=x]
    coordinates {(1,100)(5,100)(10,100)(20,100)(30,100)(40,100)(50,100)};\legend{Imputation rate (\%), Accuracy (\%)}
    \addlegendimage{orange, mark=o}
\addlegendentry{Runtime (s)}
\nextgroupplot[title = {\emph{(b) GlobalStore}},   
    xmin=0, xmax=50,
    ymin=0, ymax=100,
    xtick={0,10,20,30,40,50},
    ytick={0,20,40,60,80,100},
    legend pos=north west,
    ymajorgrids=true,
    grid style=dashed,]
\addplot[color=blue,mark=square]
    coordinates {(1,95.20)(5,95.11)(10,94.96)(20,94.67)(30,94.15)(40,93.62)(50,92.66)};
\addplot[color=red,mark=x]
coordinates {(1,97.14)(5,97.30)(10,97.08)(20,97.17)(30,97.48)(40,97.4)(50,97.04)};
    \nextgroupplot[title = {\emph{(c) RegionalSales}},  
    xmin=0, xmax=50,
    ymin=0, ymax=100,
    xtick={0,10,20,30,40,50},
    ytick={0,20,40,60,80,100},
    legend pos=north west,
    ymajorgrids=true,
    grid style=dashed,]
\addplot[color=blue,mark=square]
    coordinates {(1,88.33)(5,84.17)(10,81.63)(20,77.71)(30,73.15)(40,66.67)(50,61.73)};
\addplot[color=red,mark=x]
    coordinates {(1,92.68)(5,92.04)(10,92.76)(20,92.87)(30,90.29)(40,91.74)(50,91.42)};
\nextgroupplot[
    xmin=0, xmax=50,
    ymin=0, ymax=100,
    xtick={0,10,20,30,40,50},
    ytick={0,20,40,60,80,100},
    legend pos=north west,
    ymajorgrids=true,
    grid style=dashed,]
\addplot[color=orange,mark=o]
    coordinates {(1,1.5)(5,7.51)(10,15.61)(20,30.1)(30,43.93)(40,59)(50,76.33)};
\nextgroupplot[
    xlabel={Missing rate (\%)},   
    xmin=0, xmax=50,
    ymin=0, ymax=300,
    xtick={0,10,20,30,40,50},
    ytick={0,50,100,150,200,250,300},
    legend pos=north west,
    ymajorgrids=true,
    grid style=dashed,]
\addplot[color=orange,mark=o]
    coordinates {(1,5.79)(5,29.74)(10,48.36)(20,100.79)(30,147.42)(40,199.13)(50,252.7)};
\nextgroupplot[
    xmin=0, xmax=50,
    ymin=0, ymax=1,
    xtick={0,10,20,30,40,50},
    ytick={0,0.2,0.4,0.6,0.8,1},
    legend pos=north west,
    ymajorgrids=true,
    grid style=dashed,]
\addplot[color=orange,mark=o]
    coordinates {(1,0.04)(5,0.14)(10,0.21)(20,0.39)(30,0.59)(40,0.68)(50,0.82)};
\end{groupplot}
\end{tikzpicture}
\caption{Intra-dimensional imputation experiment results}
\label{intraIR}
\end{figure}

In Figure \ref{intraIR}, imputation rates (X-axis) vary with respect to missing rates (Y-axis). % IL FAUT TOUJOURS DECRIRE LES FIGURES !
We observe that, for dataset \textbf{TPCH}, the imputation rate is always 100\%, while the imputation rates of the other datasets decrease when the missing rate increases. The imputation rate of \textbf{RegionalSales} is much lower than the two others. Since missing data are replaced by the tuple having the same value on a lower-granularity parameter, the imputation rate of an attribute depends on the ratio of the distinct values and the coefficient of variation of each distinct value of its lower-granilarity parameters. For example, in the dimension $Part$ of \textbf{TPCH}, imputation rate and ratio are 0.125\% and 0.027 for the second granularity level parameter, respectively; while in dimension $StoreLocation$ of \textbf{RegionalSales}, they are 62.13\% and 1.1, respectively.%\vspace{-10pt}
%$Partkey \rightarrow Brand \rightarrow Manufacture$. The imputation rate of $Manufacture$ depends on the percentage of the distinct values of $Brand$, there are 25 distinct values of $Brand$ and 20000 tuples in the dimension. So they account for only 0.125\% of the total tuple. The coefficient of variation of each distinct value number is 0.027. While for example, in the dimension $StoreLocation$ of \textbf{RegionalSales}, the distinct values of the second level parameter account for 62.13\% of the total tuple and the coefficient of variation of each distinct value number is 1.1.

\iffalse
\begin{figure}
\centering
\begin{tikzpicture}
\begin{groupplot}[
   group style={
     group size=3 by 1,
     horizontal sep=1cm},
   width=\linewidth/3,/tikz/font=\scriptsize]
\nextgroupplot[title = {\emph{(a) TPCH}},    
    ylabel={Accuracy (\%)},
    xmin=0, xmax=50,
    ymin=0, ymax=100,
    xtick={0,10,20,30,40,50},
    ytick={0,20,40,60,80,100},
    legend pos=north west,
    ymajorgrids=true,
    grid style=dashed,]

\addplot[color=blue,mark=x]
    coordinates {(1,100)(5,100)(10,100)(20,100)(30,100)(40,100)(50,100)};

\nextgroupplot[title = {\emph{(b) GlobalStore}}, 
    xlabel={Missing rate (\%)},   
    xmin=0, xmax=50,
    ymin=0, ymax=100,
    xtick={0,10,20,30,40,50},
    ytick={0,20,40,60,80,100},
    legend pos=north west,
    ymajorgrids=true,
    grid style=dashed,]

\addplot[color=blue,mark=x]
    coordinates {(1,97.14)(5,97.30)(10,97.08)(20,97.17)(30,97.48)(40,97.4)(50,97.04)};
    
\nextgroupplot[title = {\emph{(c) RegionalSales}},  
    xmin=0, xmax=50,
    ymin=0, ymax=100,
    xtick={0,10,20,30,40,50},
    ytick={0,20,40,60,80,100},
    legend pos=north west,
    ymajorgrids=true,
    grid style=dashed,]

\addplot[color=blue,mark=x]
    coordinates {(1,92.68)(5,92.04)(10,92.76)(20,92.87)(30,90.29)(40,91.74)(50,91.42)};
\end{groupplot}

\end{tikzpicture}
\caption{Accuracy of intra-dimensional imputation}
\label{intraAC}
\end{figure}
\fi

%\vspace{-20pt}
\subsubsection{Accuracy}
We can see in Figure \ref{intraIR} that the accuracy of \textbf{TPCH} is always 100\%, while the accuracy of the two other datasets is always less than 100\%. Since our imputation method is based on the hierarchy relationships, non-strict and incomplete hierarchies may impact accuracy. By analysing the data, we find that there are non-strict hierarchies in these datasets \textbf{GlobalStore} and \textbf{RegionalSales}, i.e., in the dimension $Customer$ of \textbf{GlobalStore}, there are some tuples whose $City$ values are the same, but they belong to different $State$s. There is a similar case in dimension $StoreLocation$ of \textbf{RegionalSales}.

\iffalse
\begin{figure}
\centering
\begin{tikzpicture}
\begin{groupplot}[
   group style={
     group size=3 by 1,
     horizontal sep=1cm},
   width=\linewidth/3,/tikz/font=\scriptsize]
\nextgroupplot[title = {\emph{(a) TPCH}},    
    ylabel={Run time (s)},
    xmin=0, xmax=50,
    ymin=0, ymax=100,
    xtick={0,10,20,30,40,50},
    ytick={0,20,40,60,80,100},
    legend pos=north west,
    ymajorgrids=true,
    grid style=dashed,]

\addplot[color=blue,mark=o]
    coordinates {(1,1.5)(5,7.51)(10,15.61)(20,30.1)(30,43.93)(40,59)(50,76.33)};

\nextgroupplot[title = {\emph{(b) GlobalStore}}, 
    xlabel={Missing rate (\%)},   
    xmin=0, xmax=50,
    ymin=0, ymax=300,
    xtick={0,10,20,30,40,50},
    ytick={0,50,100,150,200,250,300},
    legend pos=north west,
    ymajorgrids=true,
    grid style=dashed,]

\addplot[color=blue,mark=o]
    coordinates {(1,5.79)(5,29.74)(10,48.36)(20,100.79)(30,147.42)(40,199.13)(50,252.7)};
    
\nextgroupplot[title = {\emph{(c) RegionalSales}},  
    xmin=0, xmax=50,
    ymin=0, ymax=1,
    xtick={0,10,20,30,40,50},
    ytick={0,0.2,0.4,0.6,0.8,1},
    legend pos=north west,
    ymajorgrids=true,
    grid style=dashed,]

\addplot[color=blue,mark=o]
    coordinates {(1,0.04)(5,0.14)(10,0.21)(20,0.39)(30,0.59)(40,0.68)(50,0.82)};

\end{groupplot}

\end{tikzpicture}
\caption{Run time of intra-dimensional imputation}
\label{IntraRT}
\end{figure}
\fi
%\vspace{-10pt}
\subsubsection{Runtime}
%The runtime of each dataset for each missing rate is shown in Figure \ref{intraIR}. We see that 
The evolution of runtime with respect to missing rate is linear (Figure~\ref{intraIR}), which is in line with the complexity of Algorithm~\ref{algo:duplicate1}, which is $O(n)$, where $n$ is the missing rate.

\subsection{Inter-dimensional Imputation Experiments}
\label{sec:inter}

We use \textbf{TPCH} and \textbf{GeoFrance} in this inter-dimensional imputation experiment. There are two \textbf{TPCH} dimensions, $Customer$ and $Supplier$, which have same geographical attributes. In \textbf{GeoFrance}, we create two dimensions and randomly divide the original data into two partitions with the same number of tuples. Then, we load the each partition into one of the dimensions. The imputation rate ranges between 48.67\% and 100\%.  The accuracy always remain at 100\% with respect to missing rate.% varying from 1\% to 50\%.
%\vspace{-10pt}

%\vspace{-20pt}
\subsubsection{Imputation Rate}
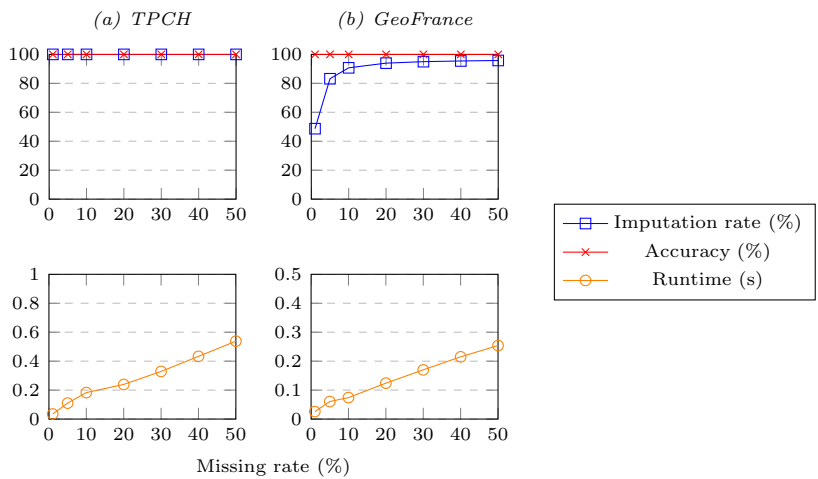
\begin{figure}
\centering
\begin{tikzpicture}
\begin{groupplot}[
   group style={
     group size=2 by 2,
     horizontal sep=1cm},
   width=\linewidth/3,/tikz/font=\scriptsize]
\nextgroupplot[title = {\emph{(a) TPCH}},   
    xmin=0, xmax=50,
    ymin=0, ymax=100,
    xtick={0,10,20,30,40,50},
    ytick={0,20,40,60,80,100},
    ymajorgrids=true,
    grid style=dashed,
    legend style={at={(2.7, -0.7)},anchor=south west}
    ]

\addplot[color=blue,mark=square]
    coordinates {(1,100)(5,100)(10,100)(20,100)(30,100)(40,100)(50,100)};
\addplot[color=red,mark=x]
    coordinates {(1,100)(5,100)(10,100)(20,100)(30,100)(40,100)(50,100)};\legend{Imputation rate (\%), Accuracy (\%)}
        \addlegendimage{orange, mark=o}
\addlegendentry{Runtime (s)}
\nextgroupplot[title = {\emph{(b) GeoFrance}}, 
    xmin=0, xmax=50,
    ymin=0, ymax=100,
    xtick={0,10,20,30,40,50},
    ytick={0,20,40,60,80,100},
    legend pos=north west,
    ymajorgrids=true,
    grid style=dashed,]

\addplot[color=blue,mark=square]
    coordinates {(1,48.67)(5,83.17)(10,90.72)(20,93.95)(30,94.97)(40,95.45)(50,95.79)};
\addplot[color=red,mark=x]
    coordinates {(1,100)(5,100)(10,100)(20,100)(30,100)(40,100)(50,100)};
\nextgroupplot[
    xmin=0, xmax=50,
    ymin=0, ymax=1,
    xtick={0,10,20,30,40,50},
    ytick={0,0.2,0.4,0.6,0.8,1},
    legend pos=north west,
    ymajorgrids=true,
    grid style=dashed,]
    
\addplot[color=orange,mark=o]
coordinates {(1,0.035)(5,0.11)(10,0.183)(20,0.239)(30,0.329)(40,0.433)(50,0.538)};

\nextgroupplot[ 
    xmin=0, xmax=50,
    ymin=0, ymax=0.5,
    xtick={0,10,20,30,40,50},
    ytick={0,0.1,0.2,0.3,0.4,0.5},
    legend pos=north west,
    ymajorgrids=true,
    grid style=dashed,]

\addplot[color=orange,mark=o]
    coordinates {(1,0.026)(5,0.0605)(10,0.0742)(20,0.124)(30,0.17)(40,0.215)(50,0.254)};

\end{groupplot}
\scriptsize
\node[anchor=north] (title-x) at ($(group c1r2.south east)!0.5!(group c2r2.south west)-(0,0.415cm)$) {Missing rate (\%)};
\end{tikzpicture}
\caption{Inter-dimensional imputation experiment results}
\label{interIR}
\end{figure}

%The imputation rates are shown in Figure \ref{interIR}. 
Yet again, the imputation rate of \textbf{TPCH} %for the given missing rates 
is always 100\% (Figure~\ref{interIR}), for the same same reason as intra-dimensional imputation. \textbf{GeoFrance}'s imputation rate is low when the missing rate is very low, then it increases with the missing rate. After analysing the data, we find that there is a tuple where the values of $RegionCode$ and $RegioName$ are originally missing. The lower-granularity parameter $DepartmentCode$ being unique, $RegionCode$ and $RegionName$ missing data cannot be imputed. When the missing rate is low, the number of total missing data is low, too. Thus, the missing data in this tuple account for a large proportion of total missing data, which can explain why \textbf{GeoFrance}'s imputation rate is low when the missing rate is very low.

\begin{comment}
\begin{figure}
\centering
\begin{tikzpicture}
\begin{groupplot}[
   group style={
     group size=2 by 1,
     horizontal sep=1cm},
   width=\linewidth/3,/tikz/font=\scriptsize]
\nextgroupplot[title = {\emph{(a) TPCH}},    
    ylabel={Accuracy (\%)},
    xlabel={Missing rate (\%)},  
    xmin=0, xmax=50,
    ymin=0, ymax=100,
    xtick={0,10,20,30,40,50},
    ytick={0,20,40,60,80,100},
    legend pos=north west,
    ymajorgrids=true,
    grid style=dashed,]

\addplot[color=blue,mark=x]
    coordinates {(1,100)(5,100)(10,100)(20,100)(30,100)(40,100)(50,100)};

\nextgroupplot[title = {\emph{(b) GeoFrance}}, 
    xmin=0, xmax=50,
    ymin=0, ymax=100,
    xtick={0,10,20,30,40,50},
    ytick={0,20,40,60,80,100},
    legend pos=north west,
    ymajorgrids=true,
    grid style=dashed,]

\addplot[color=blue,mark=x]
    coordinates {(1,100)(5,100)(10,100)(20,100)(30,100)(40,100)(50,100)};
\end{groupplot}

\end{tikzpicture}
\caption{Accuracy of inter-dimensional imputation}
\label{interAC}
\end{figure}
\end{comment}

%\vspace{-10pt}
\subsubsection{Accuracy}
There is no incomplete nor non-strict hierarchy in \textbf{TPCH}'s and \textbf{GeoFrance}'s DWs. Hence, the accuracy for these two datasets is always 100\% (Figure~\ref{interIR}).

%\vspace{-10pt}
\subsubsection{Runtime}
%The runtime of each dataset for each missing rate is shown in Figure \ref{interIR}. The change of the run time with respect to the missing rate is linear since the complexity of the algorithm is O(n), where n is the missing rate.
Again, the evolution of runtime with respect to missing rate is linear (Figure~\ref{interIR}), which is in line with the complexity of Algorithm~\ref{algo:duplicate2}, which is $O(n)$, where $n$ is the missing rate. 
\iffalse
\begin{figure}
\centering
\begin{tikzpicture}
\begin{groupplot}[
   group style={
     group size=2 by 1,
     horizontal sep=1cm},
   width=\linewidth/3,/tikz/font=\scriptsize]
\nextgroupplot[title = {\emph{(a) TPCH}},    
    ylabel={Run time (s)},
    xmin=0, xmax=50,
    ymin=0, ymax=1,
    xtick={0,10,20,30,40,50},
    ytick={0,0.2,0.4,0.6,0.8,1},
    legend pos=north west,
    ymajorgrids=true,
    grid style=dashed,]
    
\addplot[color=blue,mark=o]
coordinates {(1,0.035)(5,0.11)(10,0.183)(20,0.239)(30,0.329)(40,0.433)(50,0.538)};

\nextgroupplot[title = {\emph{(b) GeoFrance}},  
    xmin=0, xmax=50,
    ymin=0, ymax=0.5,
    xtick={0,10,20,30,40,50},
    ytick={0,0.1,0.2,0.3,0.4,0.5},
    legend pos=north west,
    ymajorgrids=true,
    grid style=dashed,]

\addplot[color=blue,mark=o]
    coordinates {(1,0.026)(5,0.0605)(10,0.0742)(20,0.124)(30,0.17)(40,0.215)(50,0.254)};
\end{groupplot}

\end{tikzpicture}
\caption{Run time of inter-dimensional imputation}
\label{InterRT}
\end{figure}
\fi
\section{Conclusion and future work}
\label{sec:conclu}

In this article, we propose an internal data imputation method for  dimensional missing data in DWs. Our method is based on the existing data found in both intra- and inter-dimensional relationships. We take in charge the imputation of both parameters and weak attributes. The solutions are formalized as algorithms and are actually implemented. Our method is validated by a series of experiments with the different percentages of missing data of the different attributes. However, not all missing data can be completed by the existing data, thus in the future we will also combine our method with web-based methods to achieve a better imputation

\section*{Acknowledgement}
This research is funded by the French National Research Agency (ANR), project ANR-19-CE23-0005 BI4people (Business Intelligence for the people).

\bibliographystyle{abbrv}
\bibliography{biblio}
\end{document}